\documentclass[useAMS,usenatbib]{mn2e}
\usepackage[cp866]{inputenc}
\usepackage[english]{babel}
\usepackage{graphicx}
\usepackage{amsmath}
\usepackage{amssymb}
\usepackage{epsf}
\usepackage{bm}

\newcommand{\rate}{\mbox{${\rm erg~cm^{-3}~s^{-1}}$}}
\newcommand{\wrate}{\mbox{${\rm erg~s^{-1}}$}}
\newcommand{\gcc}{\mbox{${\rm g~cm^{-3}}$}}
\newcommand{\beq}{\begin{equation}}
\newcommand{\eeq}{\end{equation}}
\title[Heating and cooling of magnetars with accreted envelopes]
{Heating and cooling of magnetars with accreted envelopes}

\author[A.~D.~Kaminker et al.]
{ A.~D.~Kaminker$^{1}$,   
     A.~Y.~Potekhin$^{1,2,3}$,
     D.~G.~Yakovlev$^{1}$,     
	 and 	 
	 G.~Chabrier$^{2}$ \\	 
$^{1}$Ioffe Physical Technical Institute,
Politekhnicheskaya 26, 194021 Saint-Petersburg, Russia\\
$^{2}$Ecole Normale Super{\'e}rieure de Lyon, CRAL (UMR 5574
CNRS), 46 all{\'e}e d'Italie, 69364 Lyon, France\\
$^3$Isaac Newton Institute of Chile, St.~Petersburg Branch, Russia
}
\begin{document}
\date{Accepted 2009 February 24. Received 2009 February 21; 
in original form 2009 January 9}

\pagerange{\pageref{firstpage}--\pageref{lastpage}} \pubyear{2009}

\maketitle
\label{firstpage}

\begin{abstract}
We study the thermal structure 
and evolution of magnetars
as cooling neutron stars with a 
phenomenological
heat source in an 
internal layer.
We focus on the effect of 
magnetized ($B \ga 10^{14}$~G) 
non-accreted and accreted outermost envelopes composed of
different elements,
from iron 
to hydrogen or helium. 
We discuss a combined effect of 
thermal conduction and neutrino emission 
in the outer neutron star crust
and calculate the 
cooling of magnetars with a dipole magnetic 
field  
for various locations of the heat layer,
heat rates and magnetic field strengths.
Combined effects
of strong magnetic fields and 
light-element composition simplify the interpretation
of magnetars in our model:
these effects allow one to
interpret observations assuming less extreme (therefore, more
realistic) heating.
Massive magnetars, with fast neutrino cooling in their cores,
can have higher thermal surface luminosity. 
\end{abstract}
 

\section{Introduction}
\label{introduction}

We continue theoretical studies 
(\citealt{kypssg06}, 
hereafter Paper~I; also see \citealt{kgyg06,kypssg07})
of persistent thermal activity 
of magnetar candidates -- compact X-ray sources
which include soft gamma repeaters (SGRs)
and anomalous X-ray pulsars (AXPs).
The magnetars are thought to be warm, isolated, 
slowly rotating neutron stars of age
$t \lesssim 10^5$~yr with 
superstrong magnetic fields
$B \gtrsim 10^{14}$~G (see, e.g., 
\citealt{wt06},
for a review).  
Following many authors 
(e.g., \citealt*{cgp00,thompson01,plmg07}),     
we assume that the high level of magnetar X-ray emission 
is supported by 
the release of the magnetic energy in their interiors.
Although this assumption is widespread,
there are alternative models 
(e.g., \citealt*{chn00,alp01,eaeec07,tb05,bt07}).

In Paper~I 
we studied the thermal evolution
of magnetars as cooling isolated neutron stars
with a phenomenological heat source  
in a spherical internal layer. We analyzed 
the location and power of the 
source and compared our calculations 
with observations
of SGRs and AXPs. 
We showed
that the heat source should be located at densities 
$\rho \lesssim 4 \times 10^{11}$~\gcc,
and the heating rate should be 
$\sim 10^{37}$~\wrate\ to be 
consistent with the observational data
and with the energy 
budget of isolated neutron stars. 
A deeper location of the heat source
would be extremely inefficient to power the
surface photon emission, because 
the heat would be carried away 
by neutrinos. 
 
Here,
we refine the model of the magnetar
heat-blanketing envelope. 
In this envelope, 
the effects of superstrong magnetic fields
are especially important.
We analyze the blanketing envelope 
consisting not only of iron (Fe) but also  
of light elements (H or He) which can be provided by
accretion at the early stage of magnetar evolution. 
Chemical composition  
and strong magnetic fields 
do
affect  
thermal conduction 
in the blanketing envelope and 
the thermal structure of
magnetars. 
In addition to the results of Paper~I 
(and those of \citealt{py01} and \citealt{potekhinetal03}), 
we 
take into account neutrino
energy losses in the outer crust of the 
neutron star, which can also be important 
\citep*{pcy07}.
 
\section{Observations}
\label{observ}

For the observational basis, we take
seven sources: three SGRs and 
four AXPs listed in Table~\ref{tab:observ}.
We present their ages $t$, effective surface
temperatures $T_\mathrm{s}^\infty$ (redshifted for
a distant observer), 
and redshifted thermal luminosities $L_\mathrm{s}^\infty$. 
The estimates of 
spin-down ages $t$, 
blackbody effective surface 
temperatures $T_{\rm s}^\infty$,
and non-absorbed thermal fluxes 
are taken from the SGR/AXP online Catalog
maintained by the McGill Pulsar Group.%
\footnote{http://www.physics.mcgill.ca/{$\,^\sim$}pulsar/magnetar/main.html}
All references 
in Table~\ref{tab:observ}, except for
\citealt{rp97} and \citealt*{tem08},
are taken from 
that
Catalog. 
We do not include SGR 1627--41,
the faint X-ray pulsar 
CXO J164710.2--455216
(e.g., \citealt{mcc06})  
and the unconfirmed AXP candidate 
AX J1845.0--0258 
(e.g., \citealt{tkgg06}), 
because their ages $t$ 
are unknown
(as well as the thermal luminosity $L_{\rm s}^\infty$
of SGR 1627--41).
We also do not include 
two
AXPs,
XTE J1810--197 and 1E 1048.1--5937.
Their high pulsed fractions 
($\gtrsim 40\%$, e.g., \citealt{wt06})
indicate that their flux  
comes from a small fraction of
the surface, incompatible with the model considered here.
In addition, we have excluded
the AXP 4U 0142+61 whose observed thermal emission
can be attributed to a circumstellar dusty disk, rather than
to a neutron star surface
\citep{dk06}.

\renewcommand{\arraystretch}{1.2}
\begin{table*}
\caption[]{Observational limits on ages $t$, effective
surface temperatures $T_{\rm s}^\infty$,
and blackbody luminosities 
$L_{\rm s}^\infty$ of magnetars 
} 
\label{tab:observ}
\begin{center}
\begin{tabular}{ c l l l 
l c }
\hline
\hline
N & Source & $t$  & $T_{\rm s}^\infty$ & 
$\lg L_{\rm s}^\infty$ & Refs.$^{e)}$ \\
  &  & kyr & MK & 
  erg~s$^{-1}$ &  \\
\hline
1 & SGR 1806--20  & 0.22  &  7.5${\pm 0.8}$ &  
35.1\  $\pm 0.4$  &  
M00,\ M05 \\
2 & SGR 1900+14  &  1.1  & 5.0${\pm 0.6}$ & 
34.9\  $\pm 0.5$ &   
W01  \\
3 & 1E 1841--045 & 2.0~$^{a)}$  & 5.1${\pm 0.2}$ & 
35.15\ $\pm 0.15$ &  
VG97,\ M03  \\
4 & SGR 0526--66 & 5.0~$^{a)}$  & 6.2${\pm 0.7}$  & 
34.8\  $\pm 0.4$ &   
K03  \\
5 & CXOU J010043.1--721134 & 6.8 & 3.5$\pm 0.2$~$^{b)}$  & 
35.4\ $\pm 0.2$~$^{c)}$ &  T08 \\
6 & 1RXS J170849.0--400910 & 9.0  & 5.3$\pm 0.1$ & 
34.6\ $\pm 0.1$ &  R05  \\  
7 & 1E 2259+586 &  19~$^{a)}$ &  4.77$\pm 0.05$ & 
34.4\ $\pm 0.2$~$^{d)}$  &   
RP97,\ W04 \\
\hline
\end{tabular}
\end{center}
\small{
\begin{flushleft}
$^{a)}$ Ages of SNRs (see text for references) \\
$^{b)}$ The soft component of the double 
blackbody
({\it BB}) 
spectral model (see text for details) 
\\   
$^{c)}$ 
Total {\it BB} luminosity of both
components of the double spectral model \\ 
$^{d)}$ 
From Eq.~(\ref{Ls}) with 
$D=3.0 \pm 0.5$~kpc, the flux 
$f_{\rm thX}$ 
in the 2--10 keV energy band, and $\alpha_{\rm X} \sim 3.8$ 
\\ 
$^{e)}$ 
M00 -- \citet{mcft00}; 
M05 -- \citet{mte05}; 
W01 -- \citet{wkg01}; 
VG97 -- \citet{vg97}; 
M03 -- \citet{mskk03}; 
K03 -- \citet{kkm03}; 
T08 -- \citet{tem08}; 
R05 -- \citet{roz05}; 
RP97 -- \citet{rp97}; 
W04 -- \citet{wkt04} 
\\
\end{flushleft}
}
\end{table*}
 
\begin{figure}
\epsfysize=80mm
\epsffile{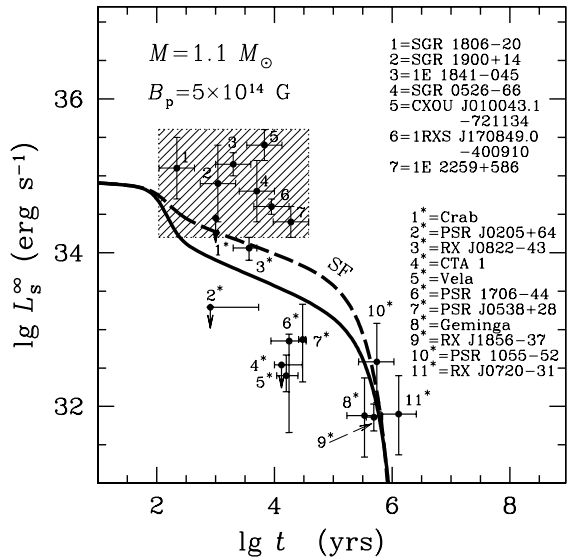}
\caption{
Observational magnetar box (the shaded rectangle) 
of blackbody luminosity 
limits, $L_{\rm s}^\infty$ 
(see text and Table~\ref{tab:observ}), 
of seven magnetars, supplemented by 
observational limits on $L_{\rm s}^\infty$ 
for eleven isolated neutron stars 
(marked by asterisks). 
The data are compared with theoretical
cooling curves $L_{\rm s}^\infty(t)$ 
for a 1.1\,$M_\odot$ 
neutron star with 
the dipole magnetic field
$B_\mathrm{p}=5 \times 10^{14}$~G (at the magnetic poles)
and no internal heating, either without superfluidity (the solid
line) or with strong proton superfluidity in the core
(the dashed line SF).
}
\label{fig1}
\end{figure}

In Fig.\ \ref{fig1} we plot 
the blackbody surface
luminosity $L_{\rm s}^\infty$  
of the selected sources 
versus their ages. 
The current data are 
uncertain and our cooling models
are too simplified to explain every
source by its own cooling model.
Instead, we will 
interpret
magnetars as cooling neutron stars belonging
to the ``magnetar box,'' the shaded rectangle
in Fig.\ \ref{fig1}\
(which reflects an average persistent thermal emission 
from magnetars, excluding bursting states). 

The thermal luminosity limits $L_\mathrm{s}^\infty$
can be obtained as
\begin{equation} 
 L_\mathrm{s}^\infty=4\pi D^2 f_\mathrm{th X} \alpha_{\rm X}.
\label{Ls}
\end{equation} 
Here, $f_\mathrm{th X}$ is a non-absorbed 
thermal flux detected from a source 
(in 
a certain
$E_1$--$E_2$
X-ray energy band), 
$D$ is a distance to the source and
\begin{equation}
 \alpha_{\rm X}=\frac{\pi^4 }{ 15} \, 
 \left( \int_{x_1}^{x_2} \frac{{\rm d}x\ x^3\ }{ \exp(x) - 1} \right)^{-1}
\label{bolomtric}
\end{equation}
is the bolometric correction,
with $x_{1,2}=E_{1,2}/k_{\rm B}T$.
In particular,
for the 2--10~keV band,
we have $\alpha_{\rm X}\sim 2$--4.  
Thermal fluxes $f_\mathrm{th X}$ 
should be inferred from observations 
with account for 
the fractions 
\textit{BB/(PL+BB)} of the blackbody ({\it BB}) 
components in the 
appropriate power law plus blackbody 
(\textit{PL+BB}) 
spectral fits (references are listed
in Table \ref{tab:observ}).
The same fits provide the effective surface
temperatures $T_\mathrm{s}^\infty$ and the
apparent radii $R_\mathit{BB}^\infty$ of emitting regions.
The radii are defined in such a way that
\begin{equation}
   L_\mathrm{s}^\infty = 4 \pi \sigma 
   (R_\mathit{BB}^\infty)^2 (T_{\rm s}^{\infty})^4,
\label{Luma}
\end{equation}
where $\sigma$ is the Stefan-Boltzmann constant.
The radii $R_\mathit{BB}^\infty$ are typically smaller than 
the
expected neutron star radii indicating that thermal
emission can originate from some fraction of a neutron star
surface. For instance, $R_\mathit{BB}^\infty \approx$2.4 and
5.5 km, for SGR 1806--20 and 1E~1841--045, respectively.  
Although 
the actual surface temperature may strongly vary 
within
the emission region, 
spectral fits give a single (surface averaged)
$T_\mathrm{s}^\infty$ value.
The luminosities $L_\mathrm{s}^\infty$
in Table \ref{tab:observ}
and Fig.\ \ref{fig1} are mainly obtained from Eq.~(\ref{Luma}) 
using the values of $T_\mathrm{s}^\infty$ and $R_\mathit{BB}^\infty$
presented in cited papers.

CXOU J010043.1--721134 
is 
the
only source from our collection, 
whose cumulative spectrum 
cannot be 
fit with a power-low plus blackbody model 
\citep{tem08}.
\citet{tem08} 
fitted it by a sum of two blackbody
components. The blackbody temperature
of the softer component 
is given
in Table~\ref{tab:observ}. 
The radius of the corresponding 
emission region, 
$R_{BB}^\infty = 12.1^{+2.1}_{-1.4}$~km,
is comparable with the theoretical 
neutron star radius, although
the pulsed fraction in the 0.2--6~keV 
energy range is rather high, $32 \pm 3 \%$.  
The harder component corresponds to
$T_{{\rm s}2}^\infty = 7.89\ ^{+1.05}_{-0.81}$~MK
and $R_\mathit{BB2}^\infty=1.7\ ^{+0.6}_{-0.5}$~km, 
meaning probably
a hot spot on the neutron star surface.
We define the thermal luminosity $L_\mathrm{s}^\infty$
of this source as
a sum of thermal luminosities of both 
blackbody components.  

Radiation from five of the 
seven selected sources 
has the overall pulsed fraction
$\lesssim 20\%$; 
the pulsed fraction 
for three of them 
is $\lesssim 10\%$ (e.g., \citealt{wt06}). 
This indicates 
that the thermal radiation can be 
emitted from a substantial part of the surface
(although the pulsed fraction is lowered by
the gravitational bending of light rays; e.g.,
\citealt{pavlovzavlin00} and references therein).
Two other
magnetars from Table \ref{tab:observ}, CXOU J010043.1--721134 
and AXP 1RXS J170849.0--400910,
have pulse fraction 
$\lesssim 40 \%$
in the 0.5--2.0~keV band
(e.g., \citealt{roz05}).

The majority of magnetar ages listed in Table~\ref{tab:observ} are 
characteristic spin down ages.
For three sources,
we adopt the ages of their host 
supernova remnants (SNRs):   
$t \sim 5$~kyr for
SGR 0526--66 in SNR N49 
(\citealt{kkm03}, also see \citealt{vblr92});
$t \sim 2$~kyr for 1E 1841--45 in SNR Kes 73 
\citep{vg97}; 
and $t \sim 19$~kyr
for 1E 2259+586 in SNR CTB 109 \citep{rp97}.  
To specify the left and right boundaries
of the magnetar box 
in Fig.~\ref{fig1}, we introduce,
somewhat arbitrarily, the uncertainties 
by a factor 2 into the ages $t$.

In our previous work we have compared
simulations of magnetar cooling with
the data on the effective surface
temperature $T_\mathrm{s}^\infty$.
Here, 
in
contrast, we use the data on $L_{\rm s}^\infty$, which 
seem more robust. Note that explaining the data either
on $T_\mathrm{s}^\infty$ or on $L_{\rm s}^\infty$
with our cooling models is not entirely self-consistent.
According to observations of all sources, but
CXOU J010043.1--721134, thermal emission originates
from some fraction of the magnetar surface while
our cooling models give thermal radiation from
a large fraction of the surface. 
If we regarded (like in Paper I) 
the temperatures $T_\mathrm{s}^\infty$ 
given by spectral fits (see Table \ref{tab:observ})
as surface-averaged effective temperatures
and calculated  $L_{\rm s}^\infty$ 
using Eq.~(\ref{Luma})
with values of $R_{BB}^\infty$ realistic for neutron stars,
we would obtain 
noticeably larger $L_{\rm s}^\infty$ 
than those provided by the observations 
(the magnetar box would
raise in Fig.\ \ref{fig1}). 
On the contrary, 
matching the theory
with the data
on $L_{\rm s}^\infty$ (as in the present paper)
gives lower surface-averaged temperatures
$T_\mathrm{s}^\infty$ than
the temperatures inferred from spectral fits. 
 
We expect that the theory will be 
improved in the future by constructing more advanced models
of magnetars -- for instance, with highly
nonuniform sources of internal energy release. 
On the other hand,
current interpretation of 
magnetar observations is far from being perfect. 
It would be a challenge
to construct new models of thermal radiation from
strongly magnetized neutron stars and use them 
(rather than blackbody models) to interpret the data. 
In this case, by analogy with employing hydrogen atmosphere models
for describing thermal radiation from ordinary neutron stars,
we
expect to get higher $R_\mathit{BB}$
(closer to the real neutron star radius) and lower
$T_\mathrm{s}^\infty$. 
Moreover, thermal radiation emitted from a magnetar
surface can be strongly distorted by magnetospheric
effects (e.g., \citealt{lg06,rea08}). 
This can greatly complicate the problem of inferring
$T_\mathrm{s}^\infty$ and $L_\mathrm{s}^\infty$ from
the data. 
 
Fig.~\ref{fig1} shows also 
observational limits $L_\mathrm{s}^\infty$ 
for eleven ordinary isolated neutron stars.
The data are taken from 
Table~1 of \citet{kgyg06} with a few
changes described by \citet{ygkp08}.
Following \citet{slane08} and \citet{shib08}, 
we have enlarged the age range of the pulsar
J0205+6449 (in 
the
SNR 3C 58)
to its characteristic age of 5.4~kyr. 
We have excluded 
one young and warm source, 1E 1207.4--5209, because
of the problems of interpretation of its spectrum.
We use the results of \citet{hoetal07} 
for the neutron star RX J1856.5--3754. The authors 
employed the magnetic hydrogen
atmosphere model 
and obtained 
$T_{\rm s}^\infty = (4.34 \pm 0.02)\times 10^5$~K 
and the apparent neutron star radius $R^\infty \approx 17$~km
(at the 68$\%$ confidence level for the fixed distance $D=140$ pc).
Taking into account a large scatter of distance estimates
for RX J1856.5--3754
\citep{walterlattimer02,vkk07}, we have added 10\%
error bars to the latter values of $T_{\rm s}^\infty$.

Fig.~\ref{fig1} shows two typical cooling
curves $L_\mathrm{s}^\infty (t)$ for a
low-mass neutron star ($M=1.1\, M_\odot$) without 
internal heating and
with a strong dipole magnetic field
($B_\mathrm{p}=5\times 10^{14}$~G at the magnetic poles). 
The solid line is 
for a non-superfluid
star, while the dashed line SF assumes a
strong proton superfluidity in the stellar core.
This superfluidity 
suppresses neutrino emission in the core and
thereby increases $L_\mathrm{s}^\infty$
at the neutrino cooling stage 
(e.g., \citealt{yp04}).

Let us stress that the surface temperature of
these stars is highly nonuniform; the magnetic poles
are much hotter than the equator. In all figures
we plot the total bolometric luminosity 
produced by the flux integrated over the stellar surface 
(e.g., \citealt{potekhinetal03}). 
The observations of ordinary neutron stars can be
explained by the cooling theory without any
reheating (e.g., 
\citealt*{yp04, ygkp08}). 
The magnetars are much hotter (more luminous)
than the ordinary cooling neutron stars;
their observations imply that they have
additional heat sources. 
As in Paper~I, we assume that
these sources are located inside magnetars.

\section{Physics input}
\label{physics}

We have 
performed calculations with our 
cooling code \citep*{gyp01},
which simulates the thermal evolution of an
initially hot star
via neutrino emission 
from the entire stellar body and via heat conduction
to the surface and thermal photon
emission from the surface. To facilitate calculations,
the star is divided into the bulk interior and
a thin outer heat-blanketing envelope (e.g., \citealt*{gpe83})
which extends from the surface
to the layer of the density $\rho=\rho_{\rm b}\sim 10^{10}-10^{11}$~\gcc; 
its thickness is a few hundred meters. 

In the bulk interior ($\rho > \rho_{\rm b}$),
the code solves the full set of 
thermal evolution equations in the spherically symmetric approximation.
The standard version of the code 
neglects the effects of magnetic fields on thermal
conduction and neutrino emission. In the present version, 
we have included 
neutrino-pair electron synchrotron radiation in a 
magnetic field $B$, that was neglected in Paper~I.

In the blanketing envelope, the updated version of the code 
(see \citealt{pcy07}, for details) 
uses a solution of stationary
one-dimensional equations
of hydrostatic equilibrium and thermal structure
with radial heat transport, anisotropic temperature
distribution, and a dipole
magnetic field \citep{go64}. It takes into account neutrino emission and
possible heat sources 
in the envelope.
The solution, 
applied to different parts of the envelope 
with locally constant magnetic fields, 
yields temperature profiles slowly varying from one part to another. 
For a given $T=T_{\rm b}$ at $\rho=\rho_{\rm b}$, we calculate
the thermal flux emergent from different parts of the surface.
Integrating it over the surface, we obtain the total photon
luminosity $L_\mathrm{s} \equiv 4 \pi \sigma R^2 T_{\rm s}^4$,
where $T_{\rm s}$ is the effective temperature properly
averaged over the stellar surface and $R$ the circumferential
neutron star radius. Redshifting then 
for a distant observer, we
have $T_{\rm s}^\infty=T_{\rm s}\,\sqrt{1-r_{\rm g}/R}$ and
$L_\mathrm{s}^\infty=(1-r_{\rm g}/R)\,L_\mathrm{s}$
($r_\mathrm{g}=2GM/c^2$ being the Schwarzschild radius).

In the present calculations,
we set either
$\rho_{\rm b}=4\times 10^{11}$~\gcc\
or (in the majority of cases)
$\rho_{\rm b}=10^{10}$~\gcc\ %
(see Sect.~\ref{structure}).
We consider blanketing envelopes composed of ground-state or accreted matter. 
For the ground state matter, the composition is 
iron up to $\rho = 10^8$~\gcc\ 
and heavier elements (e.g., \citealt*{haenseletal07})
at higher $\rho$ (it will be called the Fe composition).
As an alternative, we have studied
fully accreted envelopes composed successively of 
H, He, C, O up to maximum $\rho$ and $T$,
where these elements can survive against pycno- or thermonuclear burning,
and then the Fe composition
(we use the same structure of the accreted envelope 
as in \citealt{potekhinetal03}).
We have also considered accreted envelope models 
with H replaced by He, but found no significant difference.    

An anisotropy of heat conduction,
produced by strong magnetic fields, 
essentially modifies the temperature distribution
in the blanketing envelope (which is
included in our calculations).
The anisotropy of heat conduction
can also create anisotropic temperature distribution 
at $\rho > \rho_{\rm b}$, particularly, in the inner crust
(that is not included in our code).
Such situations should be simulated with a two-dimensional 
cooling code, as has been 
done by 
\citet*{gkp04,gkp06,pamp06} (for stationary cases)
and most recently by \citet*{apm07,apm08,aguilieraetal09,pmg09}.
The effect strongly depends on the values of thermal
conductivity across the magnetic field. If one restricts oneself
to the thermal conductivity of strongly degenerate electrons,
the anisotropy of heat conduction (the ratio of thermal
conductivities along and across the field) in a deep and not
very hot magnetized crust can be huge (because
of the strong magnetization of electrons which
are mainly moving along magnetic field lines). However, in a hot crust
(with increasing temperature) the electron
magnetization and heat conduction anisotropy weaken. Moreover,
the conductivity of phonons (lattice vibrations 
of Coulomb crystals of atomic nuclei; 
\citealt{chugunovhaensel07}) and vibrations
of superfluid neutron liquid 
(superfluid phonons; \citealt{aguilieraetal09})
in the inner crust can be much larger than the electron
thermal conductivity across the field lines, 
washing out the temperature anisotropy
and producing a nearly isotropic temperature distribution
in the deep crust (although the electron conduction
still dominates along field lines). 
This tendency is quite visible in recent calculations
of \citet{aguilieraetal09} and \citet{pmg09} and justifies
our approach.   
  
At $T_{\rm b} \ga 10^9$~K
the neutrino emission affects
the thermal structure 
of the star. We calculate the neutrino emission 
in the crust (in the blanketing envelope and deeper) taking into account 
electron-positron pair annihilation, 
plasmon decay, 
neutrino bremsstrahlung 
in collisions of electrons 
with atomic nuclei,
and synchrotron radiation
of neutrino pairs by electrons
(e.g., \citealt{yakovlevetal01}).   
The pair annihilation is relatively unimportant
and can be neglected.
Very strong magnetic fields in the outer crust of magnetars
can modify the plasmon decay
and bremsstrahlung neutrino processes. 
Such modifications have not been
studied in detail; they can be important 
for the magnetar physics but are neglected here.
Evidently, the synchrotron emission does depend
on the magnetic field, which we take into account explicitly;
it is important at 
$B \sim 10^{14}-10^{15}$~G (cf.\ \citealt{pcy07}).

\begin{figure}
\epsfysize=60mm
\epsffile{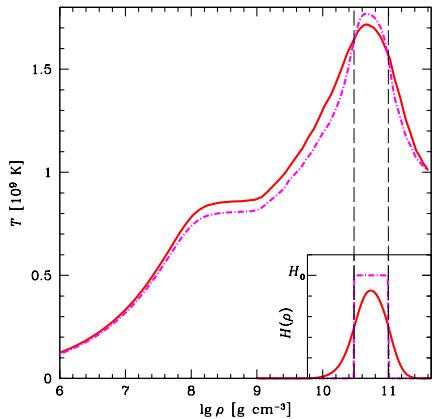}
\caption{ 
Temperature profiles along the magnetic axis of
the neutron star with $M=1.4\,M_\odot$,
$B=10^{15}$~G, $H_0=3\times10^{20}$ \rate,
and two forms of the heat source profile,
shown in the inset:
smooth (solid lines)
and piecewise-constant (dot-dashed lines). 
}
\label{fig:hform}
\end{figure}

Following Paper~I, we introduce
an internal phenomenological
heat source located in a spherical layer, $\rho_1<\rho<\rho_2$.
The heat rate $H$ [\rate] is
taken in the form
\begin{equation}
   H=H_0\,\Theta(\rho)\,\exp(-t/\tau),
\label{H}   
\end{equation}
where $H_0$ is the maximum heat intensity,
$\Theta\sim1$ in a density interval
$\rho_1<\rho<\rho_2$, and $\Theta$ vanishes outside
this interval 
(at $\rho\ll\rho_1$ or $\rho\gg\rho_2$), 
$t$ is the
star's age, and $\tau$ is the e-folding decay time of the heat
release. An exact shape of $\Theta(\rho)$ is unimportant for
$T_{\rm s}$, provided 
that we fix the total heat power
\begin{equation}
  W^\infty(t)= \int {\rm d}V\, {\rm e}^{2\Phi}\, H,
\label{W}
\end{equation}
where ${\rm d}V$ is a proper volume element and
$\Phi$ is the metric function that describes gravitational
redshift. An illustration of this independence is given in
Fig.~\ref{fig:hform}, which displays temperature distributions in
the crust for two different shapes of $\Theta(\rho)$. We see that a
change of a shape of $\Theta(\rho)$ leaves almost intact the
thermal structure at $\rho\ll\rho_1$ (and, therefore, it cannot
affect thermal radiation).

\renewcommand{\arraystretch}{1.2}
\begin{table}
\caption[]{Five positions of the heat layer, and 
the heat power $W^\infty$ for the 1.4\,$M_\odot$ star
with $H_0=3\times 10^{20}$ \rate\ and $t=1$ kyr}
\label{tab:heat}
\begin{center}
\begin{tabular}{ c c c c }
\hline
\hline
No. & $\rho_1$ (\gcc) & $ \rho_2$ (\gcc) & $W^\infty$ (erg s$^{-1}$) \\
\hline
\hline
I  & $3 \times 10^{10}$ & $10^{11}$ & $4.0 \times 10^{37}$ \\
II  & $10^{12}$ & $3 \times 10^{12}$ & $1.9 \times 10^{37}$ \\
III & $3 \times 10^{13}$ & $10^{14}$ & $1.1 \times 10^{38}$ \\
IV  & $2 \times 10^{10}$ & $6 \times 10^{10}$ & $3.3 \times 10^{37}$ \\
V  & $ 10^{11}$ & $3 \times 10^{11}$ & $4.3 \times 10^{37}$ \\
\hline
\end{tabular}
\end{center}
\small{
}
\end{table}

According to Paper I,
only the values $\tau \sim 10^4 - 10^5$~yr
can be consistent with the magnetar box 
(Sect.~\ref{observ}). Longer $\tau$
would require too much energy 
(Sect.~\ref{theory}).  
In this paper, we take $\tau=5 \times 10^4$ yr.

We employ the
same EOS in the neutron star core
as in Paper~I. It is the model 
denoted as APR~III by \citet{gusakovetal05}; it is 
based on the EOS of \citet*{apr98}.
According to this EOS,
the core consist of nucleons,
electrons, and muons. The maximum neutron star mass
is $M=1.929\,M_\odot$. The powerful direct Urca process of
neutrino emission 
\citep{lpph91}
is allowed only in the central kernels
of neutron stars with $M > 1.685\,M_\odot$ (at densities
$\rho> 1.275 \times 10^{15}$~\gcc).

We use two neutron star models, with 
$M=1.4\,M_\odot$ and $1.9\,M_\odot$.
The former is an example of a star with the
standard (not too strong) neutrino emission in the core
(the modified Urca process in a non-superfluid
star). In this case $R=12.27$~km and the central density 
is
$\rho_{\rm c}=9.280 \times 10^{14}$~\gcc.
The latter model ($R=10.95$~km, $\rho_{\rm c}=
2.050 \times 10^{15}$~\gcc) 
gives
an example of a star
whose neutrino emission is enhanced by the direct Urca process
in the inner core. 

\begin{figure*}
\epsfysize=100mm
\epsffile{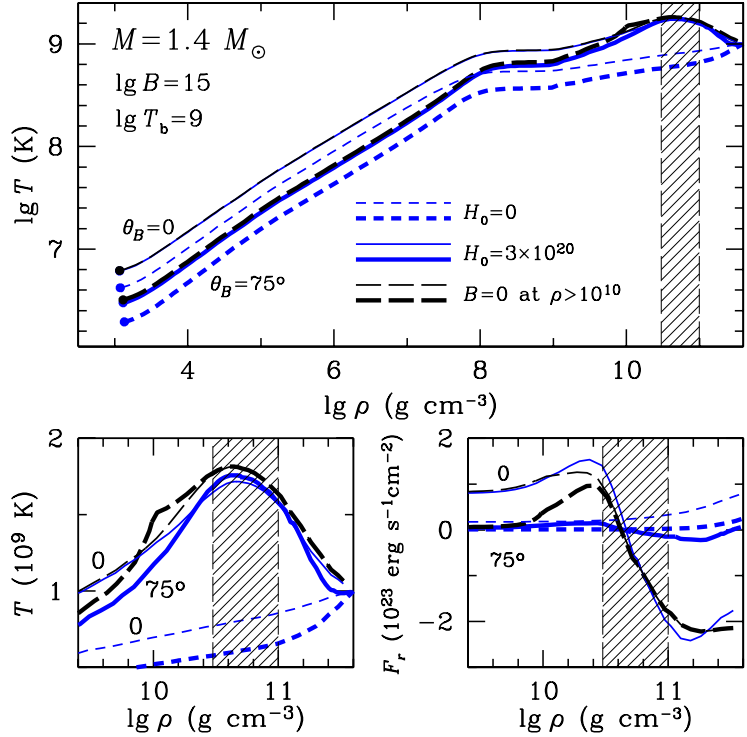}
\caption{ 
Temperature $T(\rho)$ 
and radial thermal flux $F(\rho)$ 
profiles in local parts of 
the outer crust of the 1.4$\,M_\odot$ neutron star
($^{56}$Fe envelope; $\rho_{\rm b}=4\times 10^{11}$~\gcc\ and 
fixed $T_{\rm b}=10^9$~K) with a locally  
uniform magnetic field $B=10^{15}$~G directed 
at two angles $\theta_B$ to the surface normal,
$\theta_B=0$ (thin lines) or
$\theta_B=75^\circ$ 
(thick lines). 
The position of the heat layer (I in Table~\ref{tab:heat}) is 
indicated by 
the shaded strips;
the heat intensity is
$H_0=3\times 10^{20}$~\rate;  
solid lines -- accurate calculations, 
long-dashed lines -- 
magnetic field is off 
at $\rho > 10^{10}$~\gcc;
short-dashed lines -- no heating, $H_0=0$.
{\it Top}:
entire temperature profiles at $\rho \leq \rho_{\rm b}$,
the filled dots show the radiative
surface (optical depth equals 2/3). 
{\it Bottom}: $T(\rho)$ in units of
$10^9$~K (\textit{left}) and $F(\rho)$ in units of $10^{23}$ erg~cm$^{-2}$~s$^{-1}$
(\textit{right}) in the vicinity of 
the heat layer.
}
\label{fig2}
\end{figure*}

Five examples of heat layer locations, 
$\rho_1, \, \rho_2$, 
are given in Table~\ref{tab:heat}.
Three of them (I, II, and III)  were 
used in Paper~I. 
Let us remind that the outer crust has a thickness of
a few hundred meters and a mass of $\sim 10^{-5}\,M_\odot$;
the inner crust can be as thick as $\sim1$ km and its mass
is $\sim 10^{-2}\,M_\odot$, while the core 
has radius $\sim 10$ km and contains $\sim 99\%$ of
the stellar mass. All five heat layers 
are relatively thin. The layers
I, IV, and V are located in the outer crust;
the layers II and III are at the top and bottom 
of the inner crust, respectively.
For illustration, in Table \ref{tab:heat} we present
also the heat power $W^\infty$ calculated from
Eq.\ (\ref{W}) for the five layers in the $1.4\,M_\odot$ star of age
$t=1000$ yr at $H_0=3 \times 10^{20}$~\rate.
  
\section{Thermal structure of heat-blanketing envelopes}
\label{structure}

Fig.~\ref{fig2} shows the temperature and 
radial thermal flux 
profiles in local parts of the heat blanketing Fe envelope
($\rho \leq \rho_{\rm b}=4 \times 10^{11}$~\gcc, 
$T_{\rm b}=10^9$~K) of
the $1.4\ M_\odot$ star
with a locally uniform  
magnetic field $B=10^{15}$~G
directed at
two angles to the surface normal, 
$\theta_B=0$ and $\theta_B=75^\circ$.  
One can compare the thermal structure 
of the blanketing envelope
without heating (short-dashed lines;\
e.g., \citealt{potekhinetal03,pcy07})  
and with the heat source (solid lines)
of the intensity 
$H_0=3\times 10^{20}$~\rate,
located in layer I (Table~\ref{tab:heat}).  

Long-dashed lines in 
Fig.~\ref{fig2} are obtained
by solving the full set of one-dimensional equations
for the blanketing envelope \citep{pcy07}
including the heat source 
but with the magnetic field artificially switched off
at $\rho > 10^{10}$~\gcc. These calculations simulate
the model used in Paper~I, 
where a strongly magnetized 
blanketing envelope with 
$\rho_{\rm b}=10^{10}$~\gcc\,
was matched to the interior, in which the magnetic field effects
were ignored.
We see that at different $\theta_B$ 
the long-dashed $T(\rho)$ curves
only slightly differ from the solid ones. Such a difference
is more pronounced near  
the heat layer (the bottom left panel) but becomes invisible
at lower $\rho$. 
We have obtained a significant difference
only in a narrow range
of field directions $\theta_B \approx 90^\circ$. However,
in these cases the one-dimensional (radial) model
becomes a poor approximation.
The curvature of magnetic field lines
in a more realistic model
should increase the surface temperature
in a narrow equatorial zone of width
$\lesssim\sqrt{Rh}$ (along the surface), where $h$ is the
thickness of the heat-blanketing envelope
\citep{pcy07}.
An additional increase of the surface temperature
in the above region can be provided
by ion heat conduction \citep{chugunovhaensel07}.
The temperature raise will reduce the indicated difference
between the solid and long-dashed lines.

We have obtained similar results for
local radial heat flux $F(\rho)$ shown in
the bottom-right panel of Fig.~\ref{fig2}. The flux 
changes its sign inside the heat layer. 
The flux at $\rho \gtrsim 10^{11}$~\gcc\ 
flows
into the stellar interior, where the heat 
is radiated away by neutrinos (Paper~I).
The solid and long-dashed curves $F(\rho)$ 
are also 
indistinguishable at $\rho \lesssim 10^{10}$~\gcc.
Calculations show that
the convergence of two types of the curves
is violated only at $B \gtrsim 10^{16}$~G.  

We have verified that our calculations for the blanketing envelope
with $\rho_{\rm b}=4\times 10^{11}$~\gcc\ (including the heat
layer) properly match those with 
$\rho_{\rm b}=10^{10}$~\gcc\ (with the same heat layer being outside
the blanketing envelope).
These results justify the choice of the blanketing envelope
with $\rho_{\rm b}=10^{10}$~\gcc\ in our further
calculations 
(at least with $B \lesssim 10^{15}$~G).

\begin{figure*}
\epsfysize=80mm
\epsffile{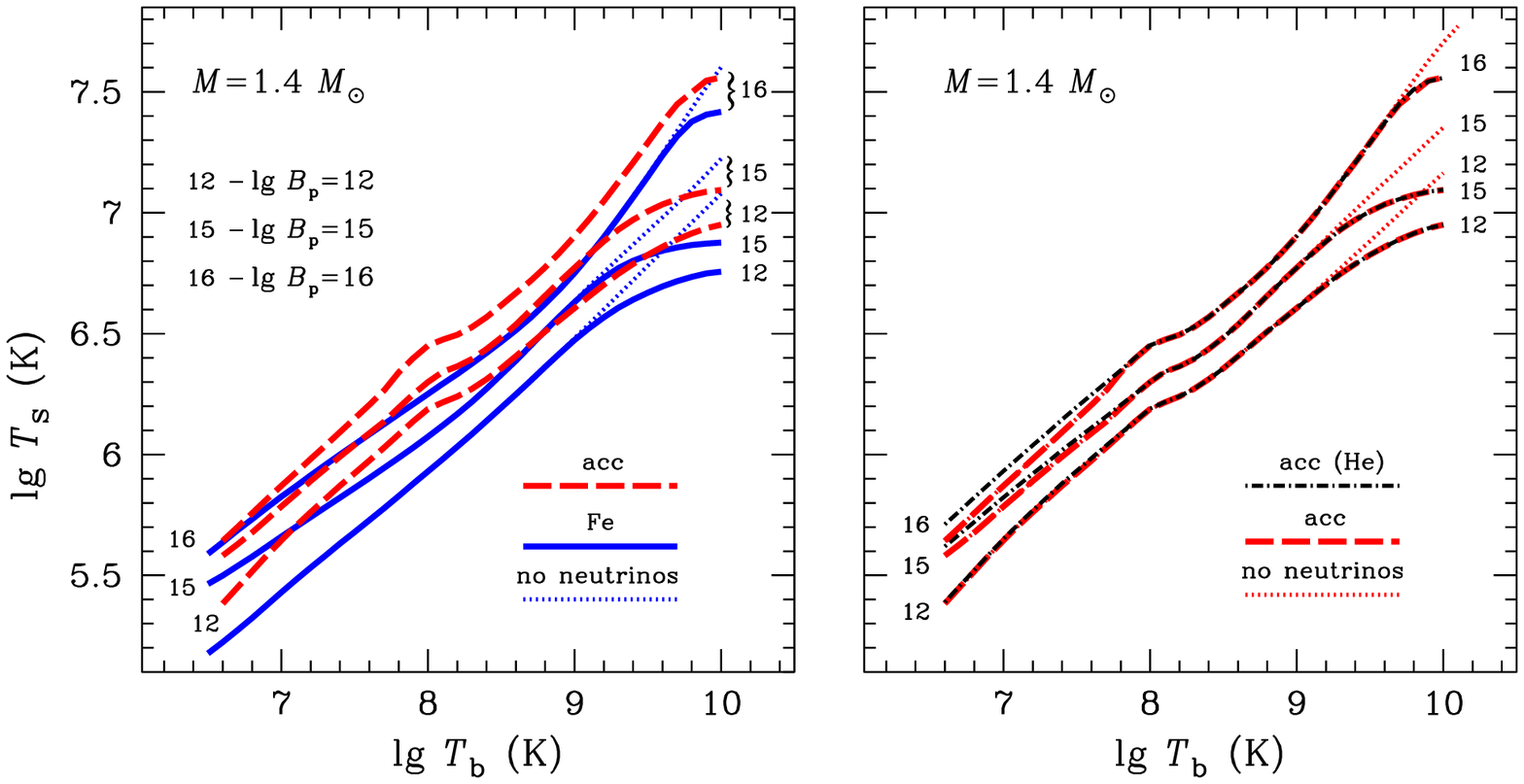}
\caption{ 
Surface temperature $T_{\rm s}$,
averaged over the surface 
of the $M=1.4 M_\odot$ neutron star, 
as a function of $T_{\rm b}$ for
the blanketing envelope with 
$\rho_{\rm b}=10^{10}$~\gcc.
Three families 
of curves  
12, 15, and 16 
correspond to the dipole magnetic fields
with 
$B_\mathrm{p}=10^{12}$, $10^{15}$, and $10^{16}$~G,
respectively. 
{\it Left}: 
Solid and dashed lines 
refer to the Fe and accreted envelopes; 
dotted lines are for the Fe envelopes  
neglecting neutrino emission. 
{\it Right}: Dashed lines -- 
standard accreted envelopes; dash-and-dot lines `acc (He)' --
the accreted envelopes with hydrogen
replaced by helium;
dotted lines -- standard accreted envelopes
neglecting neutrino emission.
} 
\label{fig5}
\end{figure*}

Fig.~\ref{fig5} shows the average surface temperature 
$T_{\rm s}$ of the $M=1.4 M_\odot$ star
versus $T_{\rm b}$. We assume 
a dipole magnetic field 
with $B_\mathrm{p}=10^{12}$,
$10^{15}$, and $10^{16}$ at the magnetic poles.
We consider our 
Fe and accreted envelopes 
as well as an accreted envelope 
with
all hydrogen
replaced by helium. 
One can see an appreciable increase of 
$T_{\rm s}$ 
with the growth of $B_\mathrm{p}$ above $10^{14}$~G
because of the 
cumulative
thermal conductivity enhancement. 
The most significant effect of the accreted 
envelopes is a systematic increase of 
$T_{\rm s}$ at any $T_{\rm b}$ 
with respect to the Fe envelope.
The effect has been studied earlier
\citep*{pcy97,py01,potekhinetal03}.
Here, we have verified that it is important  
for all magnetic fields $B_\mathrm{p}$
of our interest; it results from the   
thermal conductivity enhancement in large areas 
of the accreted envelope
near the magnetic poles.
More details on the $T_\mathrm{b}-T_\mathrm{s}$ relation
for Fe envelopes 
are given in the Appendix.

The right panel in Fig.~\ref{fig5} shows that 
replacing hydrogen by helium in the accreted envelope
at $T_{\rm b} \gtrsim 10^8$~K does not affect $T_{\rm s}$.
The insensitivity of  $T_{\rm s}$ to this replacement
for non-magnetized neutron stars 
is known \citep{pcy97}. Here, we have checked
this property for magnetars.
Only at $B_\mathrm{p} \gg 10^{14}$~G 
and relatively low temperatures $T_{\rm b} < 10^8$~K
the surface temperature $T_{\rm s}$ 
for the He envelope goes slightly higher. Generally,
such an effect is unusual for ions with higher $Z$, which
are better scatterer of electrons
(cf.\ \citealt{pcy97,potekhinetal03}). 
However,
it occurs in the superstrong field because of deeper 
localization of the radiative surface
for the He envelope,
as a result of a lower energy 
plasma-frequency cutoff in the Rosseland opacity
\citep{potekhinetal03}
for smaller $Z/A$.
Anyway at $T_{\rm b} \gtrsim 10^8$~K
replacing hydrogen by helium does not affect
the thermal insulation of the blanketing envelope.
   
Also, Fig.~\ref{fig5} shows the effects of neutrino
emission in the blanketing envelopes of different composition at 
$T_{\rm b} > 10^9$~K.
One can see that the neutrino emission
limits the growth of $T_{\rm s}$
with increasing $T_{\rm b}$ (cf.\ \citealt{pcy07}).

\begin{figure}
\epsfysize=80mm
\epsffile{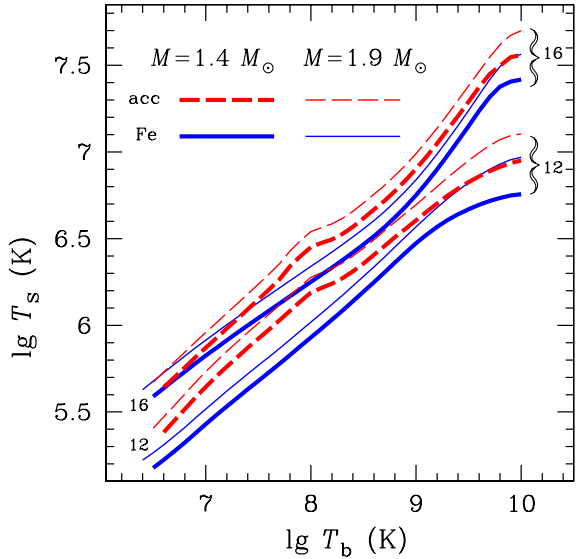}
\caption{ 
Average surface temperature $T_{\rm s}$
as a function of $T_{\rm b}$ 
for the $1.4 M_\odot$ (thick lines)
and $1.9 M_\odot$ (thin lines) stars;
solid and dashed lines
refer to the same envelopes as in Fig.~\ref{fig5}; 
labels 12 and 16 refer to the 
field strengths $B_\mathrm{p}=10^{12}$ and $10^{16}$~G. 
}
\label{fig6}
\end{figure}

Fig.~\ref{fig6} shows the dependence of  
$T_{\rm s}$ on $T_{\rm b}$ for the $1.4\,M_\odot$
and $1.9\,M_\odot$ stars 
with Fe and accreted envelopes.
In a wide range of magnetic fields, 
we obtain systematically higher 
surface temperatures $T_{\rm s}$ 
(at the same $T_{\rm b}$)
for the massive star as a result 
of smaller radius (see Sect.~\ref{physics})
and thinner
blanketing envelope 
($\Delta R_{\rm b}= R-R_{\rm b} \approx 170$~m for
$M=1.4\,M_\odot$ versus $\Delta R_{\rm b} \approx 70$~m
for $M=1.9 M_\odot$).
At a given $T_{\rm b}$, the surface temperature scales approximately
as $T_{\rm s} \propto g^{1/4}$, 
where $g\approx GM/(R^2 \sqrt{1-r_{\rm g}/R})$
is the surface gravity 
(e.g., \citealt{venturapotekhin01}, and references therein).

\section{Comparison with observations}
\label{theory}

A warm cooling neutron star with a powerful internal heating 
(Sect.~\ref{physics})
quickly (in $t \lesssim 10$ yr) 
reaches a quasi-stationary state
regulated by the heat source. 
The energy is mainly carried away by neutrinos, but some 
fraction is transported by thermal conduction to the
surface and radiated away by photons; 
the stellar interior stays 
highly non-isothermal. A thermal state
of the heat source and outer layers is almost
independent of the physics of deeper layers. 
This means the thermal 
decoupling of the heat source and outer layers 
from the deeper layers. 

In general (Paper I) heating a warm neutron
star from the core or the inner crust is inefficient
for raising $T_{\rm s}$. With the increase of
the heat power, $T_{\rm s}$ saturates because 
of the neutrino emission.
Similar saturation of $T_{\rm s}$ by the neutrino emission
was analyzed by \cite{vr91} who studied thermal response of a neutron
star to a steady-state heating. 
In Paper~I we concluded
that the heat source should be located in 
the outer crust in order to heat the surface and
be consistent with the neutron star energy budget.

\begin{figure*}
\epsfysize=80mm
\epsffile{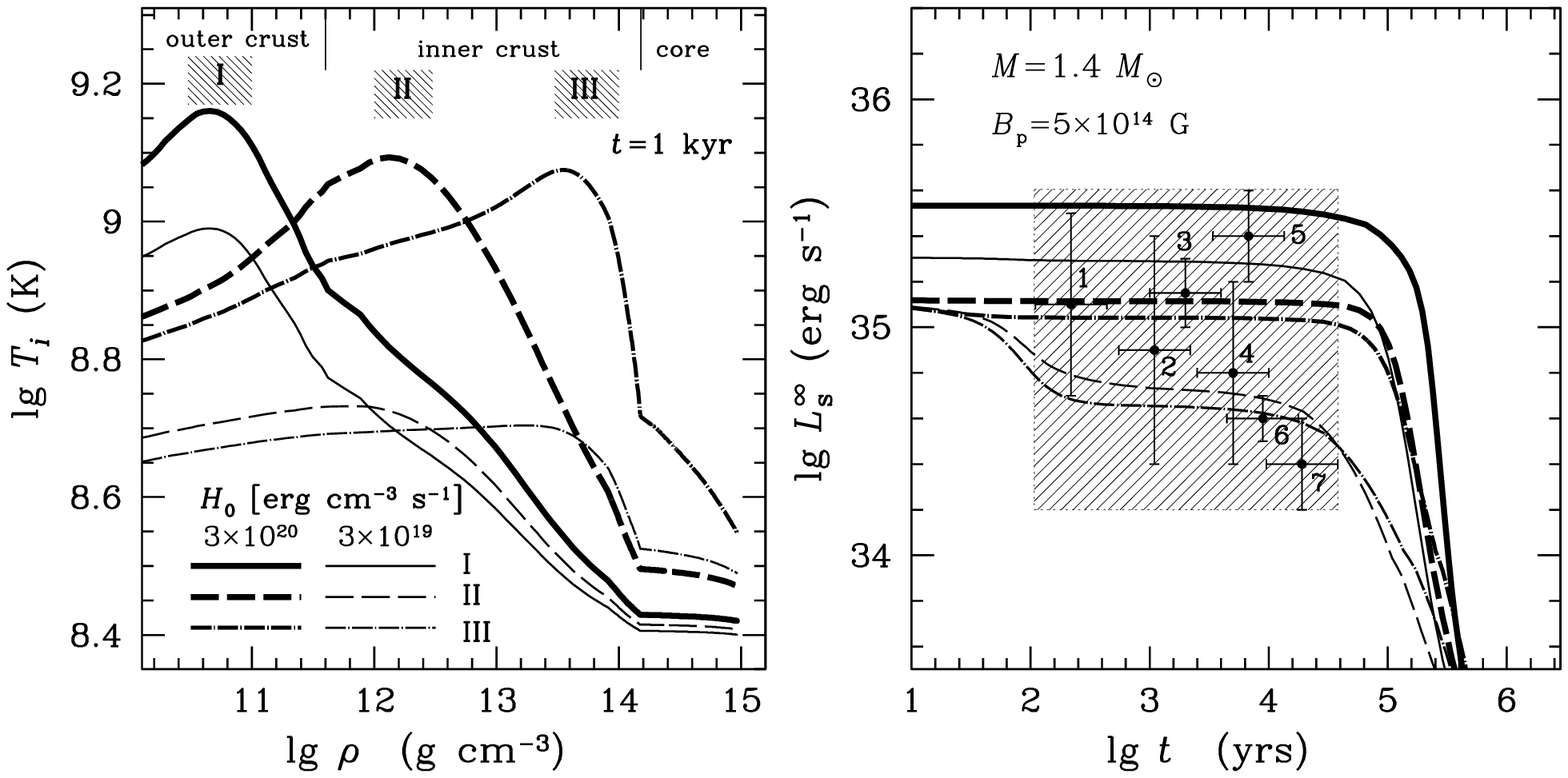}
\caption{ 
{\it Left:} Temperature dependence $T_\mathrm{i}(\rho)$ 
in the 1.4$\,M_\odot$ neutron
star of age $t=1000$ yr with three different 
positions I, II, and III (Table~\ref{tab:heat}) 
of the heat layer (hatched rectangles) and 
two levels of the heat intensity 
$H_0 = 3\times 10^{19}$~\rate\ and $3\times10^{20}$~\rate\
for the Fe blanketing envelope and
the dipole magnetic field 
with 
$B_\mathrm{p}=5 \times 10^{14}$~G at the poles.
In the upper part we indicate 
the density regions appropriate to the
outer crust, inner crust and 
the core of the star. 
{\it Right:} Cooling curves in
comparison with the magnetar box.
}
\label{fig7}
\end{figure*}
 
\begin{figure*}
\epsfysize=80mm
\epsffile{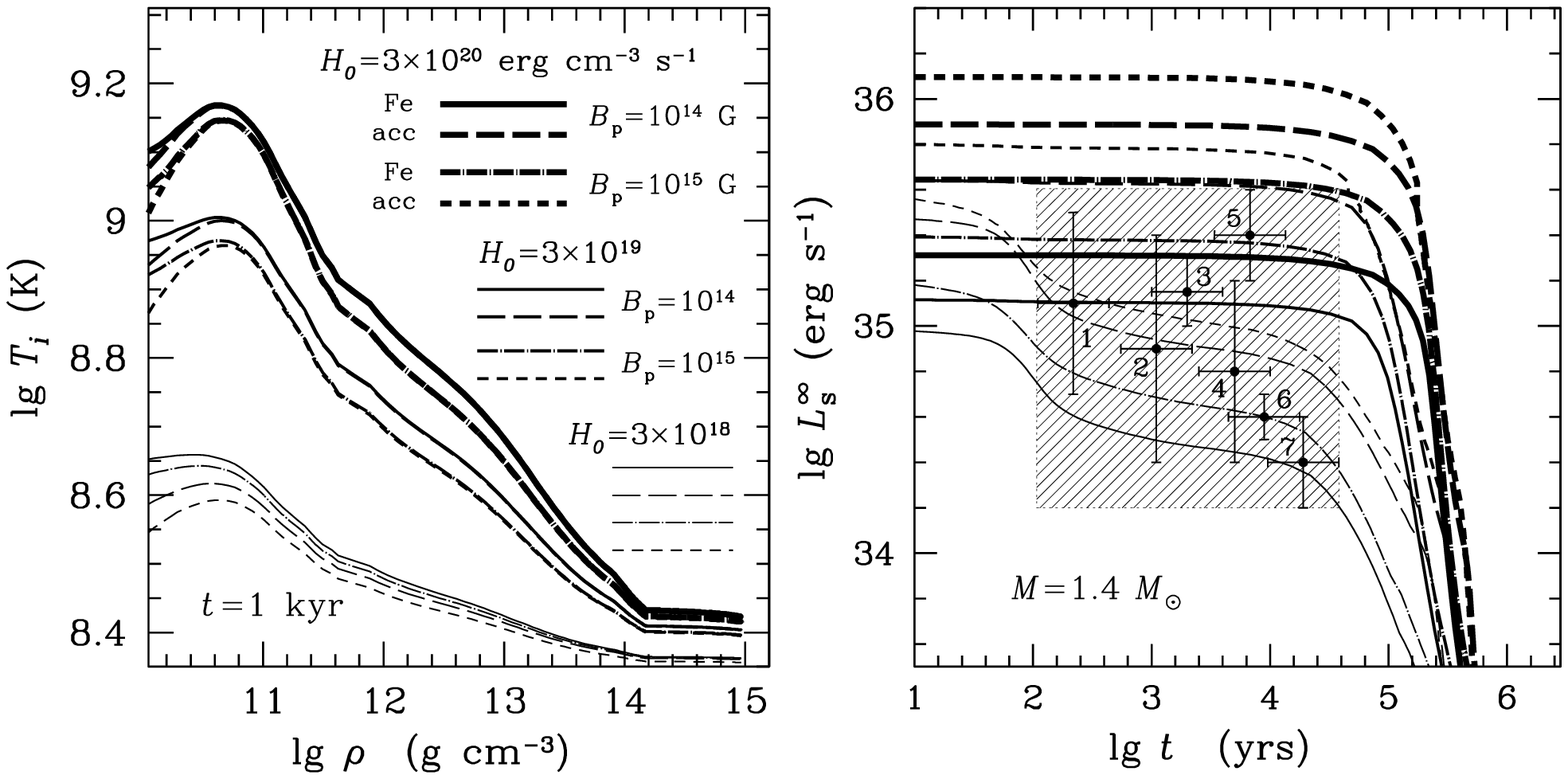}
\caption{ 
{\it Left:} Same as in the left panel of Fig.~\ref{fig7} 
but for one heat
layer I, three heat intensities,  
and two magnetic fields
($B_\mathrm{p}=10^{14}$ -- solid and long-dashed lines; 
and $B_\mathrm{p}=10^{15}$~G -- dot-dashed and short-dashed lines). 
Solid and dot-dashed lines
correspond to the Fe heat-blanketing envelope, 
dashed lines -- to the accreted envelope;
thick, intermediate, and thin lines are for the heat intensities
$H_0=3\times 10^{18}$,  $3\times 10^{19}$,  
and $3\times 10^{20}$~\rate, respectively. 
{\it Right:}  Cooling curves 
for these models.
}
\label{fig8}
\end{figure*}

\begin{figure*}
\epsfysize=80mm
\epsffile{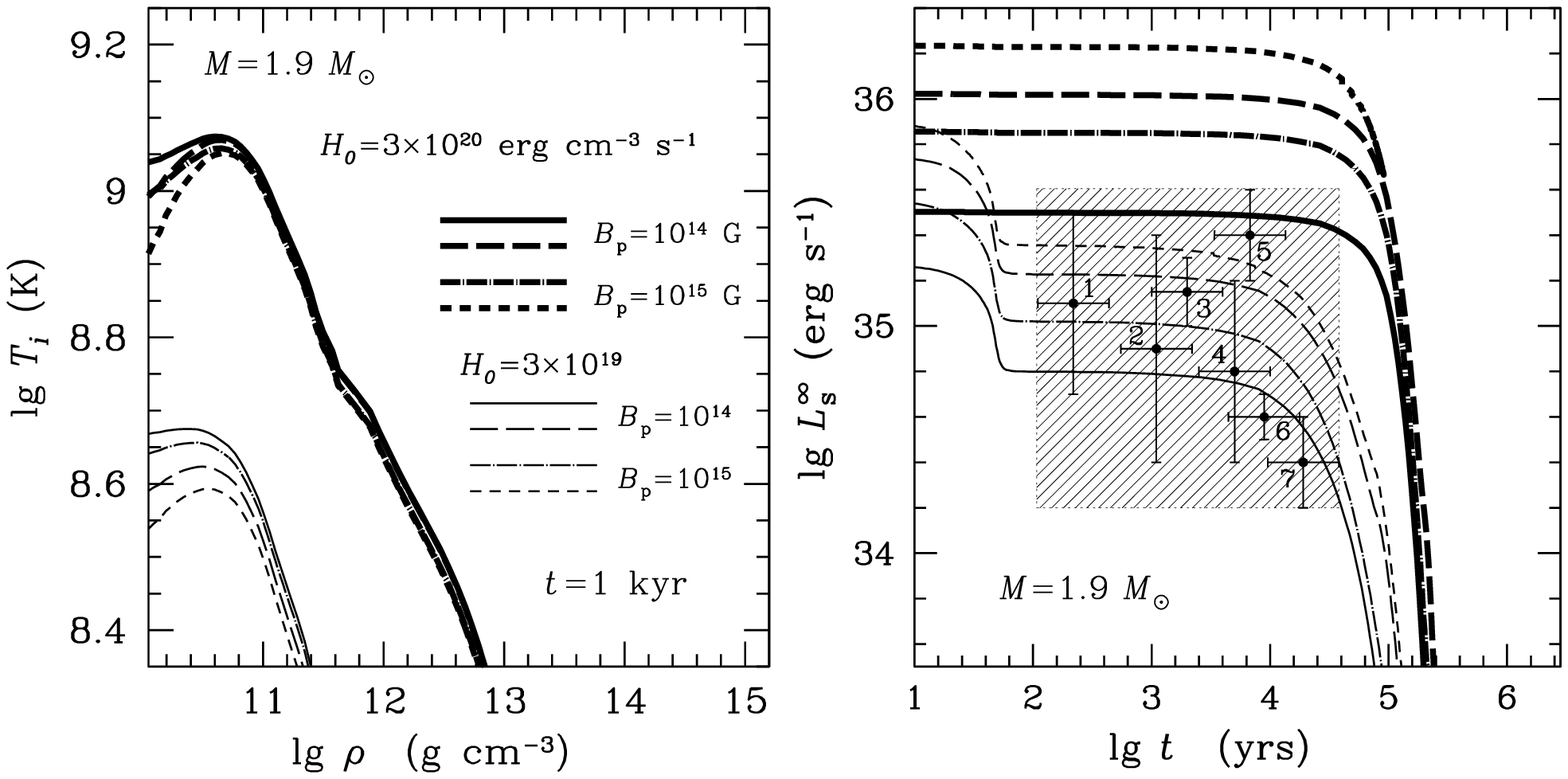}
\caption{
Same as in Fig.~\ref{fig8}
but for the 1.9\,$M_\odot$ star
and two heat intensities
$H_0=3\times 10^{19}$ (thin lines) 
and  $3\times 10^{20}$ (thick lines).  
}
\label{fig9}
\end{figure*}

\begin{figure*}
\epsfysize=80mm
\epsffile{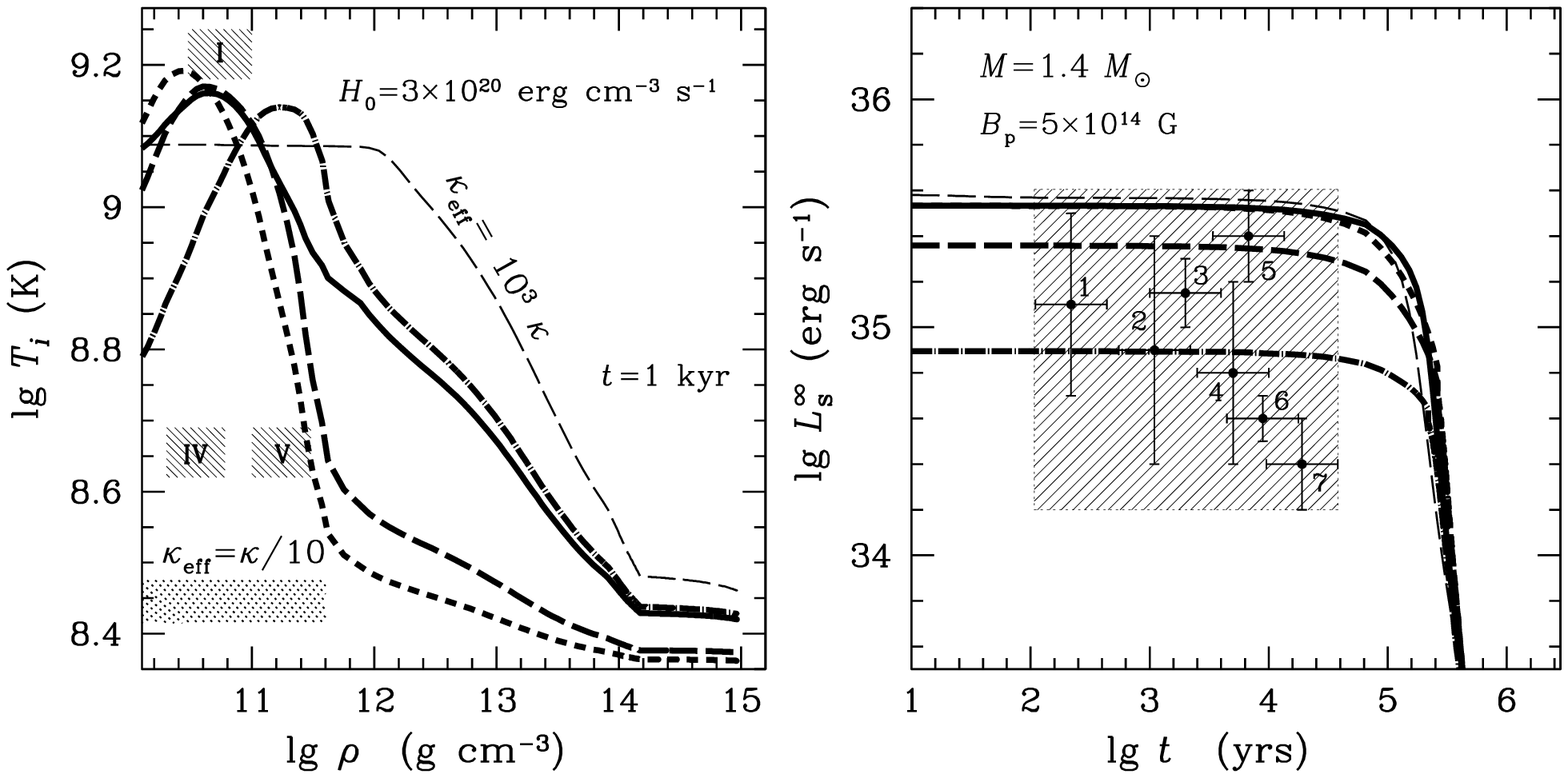}
\caption{
The effects of thermal conduction in the
outer crust on the thermal evolution of the 1.4\,$M_\odot$
star with the same magnetic field 
as in Fig.~\ref{fig7} 
at 
$H_0=3 \times 10^{20}$ \rate. 
{\it Left:} Temperature
profiles in the star at $t=$1000 yr. 
The hatched rectangles show the positions 
of the heat layers I, IV, and V, 
and the layer
(labeled as $\kappa_{\rm eff}=\kappa/10$), 
where the thermal conductivity is modified (see the text).
Solid line is the same as in Fig.~\ref{fig7}, 
thick short-dashed, long-dashed, 
and dot-dashed lines
are for the thermal conductivity reduced by a factor of
10 and for the heat layers IV, I, and V, respectively; 
thin dashed line is for the 
thermal conductivity 
enhanced by a factor of $10^3$
and the heat layer I. 
{\it Right:}
Cooling curves 
in comparison with the magnetar box. 
}
\label{fig10}
\end{figure*}

Fig.~\ref{fig7} is similar 
to Fig.~2 of Paper~I. 
It is a reference figure for 
subsequent
Figs.~\ref{fig8}--\ref{fig10}.
The left panel of Fig.~\ref{fig7} 
shows the temperature profiles $T_\mathrm{i}(\rho)$
inside the 1.4\,$M_\odot$ star 
of age $t=1000$ yr
with the dipole magnetic field
($B_\mathrm{p}=5\times 10^{14}$~G).
Here, $T_\mathrm{i}(\rho)=T(\rho) e^\Phi$ is
the internal temperature redshifted
for a distant observer, while $T(\rho)$ is the local
temperature at a given $\rho$. 
It is $T_\mathrm{i}$ that is
constant throughout thermally relaxed (isothermal)
regions of the star in 
General Relativity. The same temperature $T_\mathrm{i}(\rho)$
has been plotted in Paper I and in \citet{kypssg07} 
(denoted there as $T(\rho)$).
We consider three locations of the heat 
layer (I, II, and III in Table~\ref{tab:heat}) 
and two intensities,
$H_0=3\times10^{19}$ and $3\times10^{20}$ \rate. 
In all the cases the stellar core 
is colder 
than the crust, because of intense neutrino cooling in the core.
Pushing the heat source deeper 
into
the crust
we obtain a colder 
surface because of more efficient neutrino cooling.
 
The right panel of Fig.\ \ref{fig7} 
shows cooling curves.
Nearly horizontal parts of the curves 
at $t\lesssim \tau=5 \times 10^4$~yr
(Sect.\ \ref{physics})
and their later sharp drops  
confirm that the surface thermal luminosity
is solely maintained by internal heating. 
The initial parts ($t \lesssim 100$ yr) of 
two cooling curves for the heat layers II and III
(at $H_0 \sim 3\times 10^{19}$~\rate)
demonstrate the end of the relaxation to 
quasi-stationary thermal states.
We can
reconcile the cooling curves 
at lower heat intensity
with 
more luminous
sources in the
magnetar box 
by placing the heat source 
in the outer crust. 
To minimize the energy consumption 
(see below) we employ 
the outer heat layer I  
in Figs.\ \ref{fig8}, \ref{fig9}
and the layers I, IV, and V 
in Fig.\  \ref{fig10}.

The left panel in Fig.\ \ref{fig8} shows
the temperature profiles $T_\mathrm{i}(\rho)$ 
inside the 1.4\ $M_\odot$ star of the age $t=1000$ yr
with the heat source in layer I 
calculated for both the Fe and accreted blanketing envelopes.
For comparison, we take 
three levels 
of the heat intensity,
$H_0=3 \times 10^{18}$ (thin lines),
$H_0=3 \times 10^{19}$ (intermediate lines),
and $3 \times 10^{20}$~\rate\ (thick lines),
and two magnetic fields
($B_\mathrm{p}=10^{14}$ 
and $10^{15}$~G). The right panel 
in Fig.\ \ref{fig8}
demonstrates the appropriate cooling curves.

Under the heat-blanketing envelope 
(at $\rho > \rho_{\rm b}$),
we take into account the magnetic field effects
only by including 
the synchrotron
neutrino radiation 
and (indirectly) 
the heat source that can be provided by magnetic
fields. The synchrotron emissivity
is calculated 
by putting
$B=B_{\rm p}$. 
The neutrino synchrotron process in superstrong
magnetic fields ($10^{14}-10^{15}$~G) 
lowers the temperature 
profiles in the stellar interior; 
this effect is more pronounced for
stronger fields (cf.\ solid lines for $B_\mathrm{p}=10^{14}$ 
and dot-dashed lines for $B_\mathrm{p}=10^{15}$~G
in the left panel of Fig.~\ref{fig8}).  
The accreted matter in the 
blanketing envelope also reduces the temperature
 at 
$\rho \lesssim 10^{11}$ g~cm$^{-3}$
because of
higher heat transparency of the accreted envelope
(higher heat flux 
to the surface).

Fig.~\ref{fig8} demonstrates that 
$L_\mathrm{s}^\infty$ is mainly regulated
by the blanketing envelope. 
The combined effect
of a superstrong magnetic field and an accreted 
envelope 
appreciably
increases $L_\mathrm{s}^\infty$ (also see Fig.~\ref{fig5}).
The stronger heating ($H_0=3 \times 10^{20}$~\rate)
at $B_{\rm p}=10^{15}$~G produces too warm
magnetar envelope for any composition 
and gives larger $L_\mathrm{s}^\infty$
than required by the magnetar box. In a lower field,
$B_{\rm p}=10^{14}$~G, the accreted envelope is also too warm
but the Fe envelope is cooler 
and better 
consistent with
the magnetar box. 
The weaker (intermediate) source ($H_0=3 \times 10^{19}$~\rate)
also overheats the accreted envelopes 
at both $B_{\rm p}=10^{15}$~G and $B_{\rm p}=10^{14}$~G 
(intermediate short- and long-dashed lines);
in this case the cooling curves for Fe envelopes better match the data.
However,  only 
the weakest chosen heat intensity 
($H_0=3 \times 10^{18}$~\rate) 
is capable to cover 
   the
lower
part of 
the magnetar box.

Fig.~\ref{fig9} shows even more pronounced 
effects of the magnetar magnetic fields and accreted 
envelopes 
for the 1.9\,$M_\odot$ star  with  two levels of heat
intensities 
in the layer~I.  
The left panel 
gives the temperature profiles $T_\mathrm{i}(\rho)$
at $\rho>\rho_b=10^{10}$~\gcc. They  
are noticeably lower than 
the corresponding profiles
in  the 1.4\,$M_\odot$ star 
(because of the
direct Urca process that operates in the inner core
of the massive star).  However, 
in the case of $H_0=3 \times 10^{20}$ \rate\ 
the effect of the magnetized accreted envelope
overrides that of the rapid neutrino cooling of the massive core
(owing to thermal decoupling of the surface
from the core).
Comparing the right panels
of Figs.~\ref{fig8} and \ref{fig9}, 
we see that at 
$H_0=3 \times 10^{20}$ \rate\
and $t \lesssim 5 \times 10^4$~yr
the thermal luminosity
of the heavier $1.9\,M_\odot$ star
is higher than of the 
$1.4\,M_\odot$ star   
(because of higher
$T_{\rm s}$,
see Fig.~\ref{fig6}), making the heavier star too hot.
As discussed in Paper~I,
an intense heating in the outer crust of a massive star
can outweigh fast neutrino cooling 
in the inner core. This effect is very unusual for ordinary
cooling stars where massive stars are commonly colder than
low-mass ones (e.g., \citealt{yp04}). 

In contrast, the effects of fast neutrino cooling in magnetars
are more essential
at $H_0=3 \times 10^{19}$~\rate. Tuning 
$H_0 \sim (1-3)\times 10^{19}$~\rate\ 
and the chemical composition of the blanketing envelope,
we can reasonably well explain the magnetar box. 
Note  that  the lowest heat intensity
$H_0=3 \times 10^{19}$~\rate\
taken in Fig.~\ref{fig9} 
is ten times larger than the lowest heat intensity
$H_0=3 \times 10^{18}$~\rate\
taken in Fig.~\ref{fig8},
but the corresponding cooling curves do not strongly differ.

Earlier we \citep{kypssg07} have shown that
strong variations of the thermal conductivity
in the inner crust 
for the case of 
intense heating in layer I 
have no effect
on the surface luminosity. 
Fig.~\ref{fig10} demonstrates the sensitivity of
the thermal structure (the left panel) 
and cooling curves (the right panel) 
to artificial variations of the 
thermal conductivity in the outer crust. 
The left panel 
shows temperature profiles  
in the 1.4\ $M_\odot$  1000 yr-old star 
with the dipole magnetic field
($B_\mathrm{p}=5 \times 10^{14}$~G) 
for three locations of the heat layer 
(cases I, IV, and V in Table~\ref{tab:heat})
at $H_0=3~\times 10^{20}$~\rate;
the right panel gives 
respective cooling curves. 
The thick solid lines are the same 
as in Fig.~\ref{fig7}.
Other lines are calculated with 
the thermal conductivity modified 
in the density range 
$1.1\times 10^{10} \leq \rho \leq 4\times 10^{11}$~\gcc\
(the range is marked in the left 
panel of Fig.~\ref{fig10} by
a dot-hatched rectangle).  

Thick dashed and dot-dashed lines are 
calculated with the thermal conductivity 
reduced by a factor of 10 
($\kappa_{\rm eff}=\kappa/10$). 
They illustrate  
a possible suppression of radial heat conduction
(e.g., by a strong toroidal magnetic field
in the outer crust).   
For
the heat layers I (long-dashed lines) 
and IV (short-dashed lines) 
located closer to the bottom 
of the blanketing envelope, 
the conductivity reduction 
results in a sharper temperature drop inside 
the crust and in a cooler interior, with 
the tendency to the isothermal state.
Taking the heat 
region V (thick dot-dashed lines), shifted 
to the inner edge of the layer 
with the reduced conductivity, we obtain
qualitatively the same behavior of $T_\mathrm{i}(\rho)$, 
as in the cases II and III
in Fig.~\ref{fig7}. The thermal energy 
easier flows inside the star 
than in the case I 
with normal conduction
(solid lines). 

On the contrary, the enhanced thermal conductivity 
(thin dashed lines) 
produces a wide quasi-isothermal layer in the 
outer crust (the left panel of Fig.~\ref{fig10}) and 
a photon surface luminosity (the right panel)
that is nearly the same, as for
the normal conductivity
(the thick solid line). 
In other words,
$L_\mathrm{s}^\infty(t)$ 
is slightly sensitive to a conductivity increase.
Comparing the right panels of
Figs.~\ref{fig10} 
and \ref{fig8} (thick lines), we conclude 
that a conductivity increase
at $\rho > \rho_{\rm b}$
is incapable to rise
$L_\mathrm{s}^\infty$,
while an increase
at $\rho < \rho_{\rm b}$
can rise it.

Finally, let us discuss briefly the energy budget of magnetars.
Following Paper~I, we assume
that the maximum energy of the internal heating is
$E_{\rm max} \sim 10^{50}$~erg (which is the 
magnetic energy of the star with 
$B\sim 3\times 10^{16}$~G in the core).
Then the maximum persistent energy generation rate
is $W_{\rm max}\sim E_{\rm max}/\tau \sim 3 \times 10^{37}$
erg~s$^{-1}$. For example, let us take an 1.4$\,M_\odot$ neutron star
of age $t \ll \tau$ with the heat source in layer I.
For an intense heating with $H_0 \sim 3 \times 10^{20}$ \rate\
we obtain $W^\infty \sim W_\mathrm{max}$ (and, therefore,
$H_0$ cannot be larger). For a less intense heating with
$H_0 \sim 3 \times 10^{19}$ \rate\ we have a more relaxed condition
$W^\infty \sim 0.1\,W_\mathrm{max}$ (which would leave some energy
for bursting activity of magnetars).  

It follows from Figs.~\ref{fig7}--\ref{fig10}, 
that the heating should be sufficiently
intense to keep 
$L_\mathrm{s}^\infty$ 
on the magnetar values ($\sim 10^{35}$~erg~s$^{-1}$).
However, for realistic magnetic fields
$B \sim (2-10) \times 10^{14}$ G, the maximum allowable
heat intensity $H_0 \sim 3 \times 10^{20}$ \rate\ 
and accreted envelopes, we have
the thermal surface luminosity
$L_\mathrm{s}^\infty \sim 10^{36}$ erg~s$^{-1}$, 
noticeably higher 
than the luminosity of magnetars (Fig.~\ref{fig8}). 
For Fe envelopes and the same heat
intensity, we obtain 
$L_\mathrm{s}^\infty \gtrsim 3 \times 10^{35}$ erg~s$^{-1}$, 
consistent with the upper part of the magnetar box 
but 
giving 
the stringent energy budget 
($W^\infty \sim W_\mathrm{max}$). 
Using a weaker heat intensity $H_0 \sim 3 \times 10^{19}$ \rate\
and accreted envelopes, we obtain
still greater thermal luminosity
$L_\mathrm{s}^\infty > 4 \times 10^{35}$ erg~s$^{-1}$,
which is too high for the magnetar box 
but provides a reasonable energy budget.
Finally,   
varying  weaker  heating rate $H_0 \lesssim 10^{19}$~\rate\ 
and the chemical composition of the blanketing envelope 
(Fig.~\ref{fig8})  we have the luminosity
$L_\mathrm{s}^\infty \sim 10^{35}$ erg~s$^{-1}$,
that is consistent with the magnetar box. 

Accordingly, the presence of accreted envelopes simplifies
the explanation of magnetars as cooling neutron stars (in our
model). Our results show that we can reconcile the theory
with observations assuming the accreted envelopes and
lower heat intensities, $H_0 \sim 10^{19}$ \rate.
Note that in all the cases the efficiency of heat conversion
into the thermal radiation, $L_\mathrm{s}^\infty/W^\infty$,
is low but the accreted envelopes increase it. For instance,
assuming $H_0 \sim 3 \times 10^{19}$ \rate\ 
we have $L_\mathrm{s}^\infty/W^\infty \sim 0.01$ for Fe envelopes
and $L_\mathrm{s}^\infty/W^\infty \sim 0.1$ for accreted ones.

\section{Conclusions}
\label{conclusions}

We have analyzed the hypothesis that magnetars are isolated 
neutron stars
with $B \gtrsim 10^{14}$~G,
heated by a 
source localized in an internal layer.
We have modelled the
thermal evolution of magnetars,
taking into account that their
heat blanketing envelopes can be 
composed of light elements. Such envelopes can appear 
either due to a fallback accretion 
after a supernova explosion 
(e.g., \citealt*{ch89,ch96,cab04}),
probably with subsequent
nuclear spallation reactions \citep*{bsw92},
or due to later and more
prolonged accretion from a fossil disk 
(e.g., \citealt{chn00}; \citealt*{wck06,eaeec07,romanova08}) 
or from the interstellar medium (e.g., \citealt*{nelson_sw93,mor96}).

Compared to Paper~I, we have (i) included the effects of
accreted envelopes, and (ii) changed the strategy of reconciling the theory
with observations. We rely now
on observational limits of 
quasi-persistent thermal luminosities of magnetars; this lowers
the magnetar box
(Sect.~\ref{observ}) and relaxes theoretical constraints
on the properties of internal heat sources. 

The main conclusions are as follows:

(1) The presence of light elements 
in the outer envelope of a magnetized neutron star 
can significantly increase
the thermal conductivity 
and the
thermal stellar luminosity $L_\mathrm{s}^\infty$ (for a given
temperature $T_{\rm b}$ at the bottom of the heat blanketing envelope).
Similar conclusions have been made earlier
for ordinary cooling neutron stars
with $B \lesssim 10^{13}$~G (\citealt{pcy97,yp04}, 
and references therein) as well as for 
strongly magnetized cooling stars 
(e.g., \citealt{potekhinetal03}).

(2) The luminosity $L_\mathrm{s}^\infty$ of the star
with an accreted envelope is insensitive to replacing
all accreted hydrogen by helium
(as in ordinary neutron stars, see \citealt{pcy97}). 
In particular, these results 
can be used for taking into account rapid
nuclear burning of hydrogen 
and accumulation of helium
in the outer part
of the envelope 
(e.g., \citealt{cab04}).

(3) The combined effect of a superstrong 
magnetic field and an accreted envelope 
simplifies the interpretation of
observations of quasi-persistent thermal
radiation from magnetars using our model.
We confirm the conclusion of Paper I 
that the our most favorable models
require   
the heat source to be located in the outer crust 
(at $\rho \lesssim 4 \times 10^{11}$~\gcc). 
However, the presence of accreted envelopes 
allows us to take lower
heat intensities $H_0 \sim 10^{19}$~\rate\
and place the heat layer slightly deeper 
in the stellar interior.

(4) In accordance with Paper I, in all our successful models (with and
without accreted envelopes), 
heating of the outer crust 
produces a strongly nonuniform temperature distribution
within the star. The temperature in the heat layer
exceeds $10^9$~K, while the bottom of the crust and
the stellar core remain much colder. 
The outer crust is thermally decoupled 
from the inner layers; thermal surface emission
is rather insensitive to the properties of the inner layers
(such as the EOS, neutrino
emission, thermal conductivity, superfluidity
of baryons). 

(5) The surface thermal luminosity
is weakly affected
by variations of the thermal conductivity
in the outer crust below the heat 
blanketing envelope. Therefore, the effects of the magnetic field
on the conductivity in the
heat layer cannot
greatly change the surface luminosity.
The thermal 
surface radiation is mainly regulated by the heat source as well as by
the magnetic field
and chemical composition 
of the blanketing envelope. 
Nevertheless, our calculations can be improved 
by a more careful treatment of heat transport
in the entire magnetized outer crust, at
$\rho \gtrsim 10^{10}$ \gcc, with different
magnetic field configurations 
(e.g., \citealt{gkp06,apm07}). 
 
(6) Increasing the surface thermal emission 
of the star, which has a 
relatively high heat intensity
($H_0 \sim 10^{20}$~\rate)
and an accreted envelope, is
even more efficient if the star is massive
(and possesses, therefore, thinner and more heat transparent crust).
This effect is stronger than fast
neutrino cooling due to direct Urca process that
can be allowed in the core of a massive star.
 
(7) The presence of an accreted envelope 
can raise the  
efficiency of heat conversion into the surface radiation.
It can become as high as $\sim 10\%$
(compared to a maximum of $\sim1$\% for Fe envelopes).
This enables us to
make our models more consistent with the
total energy budget of heat sources in a neutron star. 
Now we can reduce the total energy to 
$E_{\rm tot}^\infty\sim 10^{48}-10^{49}$ erg
(instead of the previously assumed level of 
$E_{\rm tot}^\infty\sim 10^{49}-10^{50}$ erg).

Further observations as well as new models of magnetar atmospheres
are needed for more reliable interpretation of observations. 
The physics of internal heating 
is still not clear; it should be elaborated in the future.

\section*{Acknowledgments}
We are indebted to Yu.A.\ Shibanov for 
numerous discussions and critical remarks.
We thank the referee, Ulrich Geppert, for careful reading and valuable
suggestions.
AYP is grateful to Lilia Ferrario for a useful discussion
on magnetar observations.
This work was partly supported by 
the Russian Foundation for Basic Research
(grant 08-02-00837), and by the State
Program ``Leading Scientific Schools of Russian Federation''
(grant NSh 2600.2008.2).


\appendix
\section{$T_\mathrm{b}$\,--\,$T_\mathrm{s}$ relation}

\begin{table}
\caption[]{Internal temperature $T_\mathrm{b}$ [K] together with
effective surface temperature $T_\mathrm{s}$ [K] and
internal outward heat flux $F_\mathrm{b}$ [erg~s$^{-1}$~cm$^{-2}$]
(either at the magnetic pole or surface averaged, av) for a
1.4~$M_\odot$ neutron star of radius $R$=12.27~km with
the Fe envelope at 
$B_\mathrm{p}=10^{12}$~G and $\rho_\mathrm{b}=10^{10}$~\gcc.}
\label{tab:TbTs12}
\begin{center}
\begin{tabular}{ccccc}
\hline
\hline
$\lg T_\mathrm{b}$ &
 $\lg T_\mathrm{s}$ & $\lg F_\mathrm{b}$ &
 $\lg T_\mathrm{s}$ & $\lg F_\mathrm{b}$ 
  \\
         & pole  & pole & av  & av \\
\hline
  6.8  & 5.38 & 17.29  & 5.31 & 17.02  \\
  7.0  & 5.49 & 17.73  & 5.42 & 17.45  \\
  7.2  & 5.60 & 18.17  & 5.52 & 17.87  \\
  7.4  & 5.70 & 18.58  & 5.62 & 18.27  \\
  7.6  & 5.80 & 18.98  & 5.72 & 18.67  \\
  7.8  & 5.90 & 19.37  & 5.82 & 19.07  \\
  8.0  & 6.00 & 19.77  & 5.92 & 19.47  \\
  8.2  & 6.10 & 20.17  & 6.03 & 19.89  \\
  8.4  & 6.20 & 20.58  & 6.13 & 20.31  \\
  8.6  & 6.31 & 21.01  & 6.24 & 20.76  \\
  8.8  & 6.41 & 21.46  & 6.36 & 21.27  \\
  9.0  & 6.51 & 22.10  & 6.47 & 22.02  \\
  9.2  & 6.61 & 22.92  & 6.57 & 22.90  \\
  9.4  & 6.68 & 23.67  & 6.65 & 23.67  \\
  9.6  & 6.74 & 24.34  & 6.72 & 24.34  \\
  9.8  & 6.79 & 24.97  & 6.77 & 24.97  \\
\hline
\end{tabular}
\end{center}
\end{table}

\begin{table}
\caption[]{Same as in Table~\ref{tab:TbTs12} but for 
$B_\mathrm{p}=10^{15}$~G.}
\label{tab:TbTs15}
\begin{center}
\begin{tabular}{ccccc}
\hline
\hline
$\lg T_\mathrm{b}$ &
 $\lg T_\mathrm{s}$ & $\lg F_\mathrm{b}$ &
 $\lg T_\mathrm{s}$ & $\lg F_\mathrm{b}$ 
  \\
         & pole  & pole & av  & av \\  
\hline
  6.8  & 5.66 & 18.44 & 5.58 & 18.09 \\
  7.0  & 5.74 & 18.76 & 5.65 & 18.40 \\
  7.2  & 5.82 & 19.08 & 5.73 & 18.72 \\
  7.4  & 5.91 & 19.40 & 5.81 & 19.04 \\
  7.6  & 5.99 & 19.72 & 5.90 & 19.36 \\
  7.8  & 6.07 & 20.05 & 5.98 & 19.69 \\
  8.0  & 6.15 & 20.39 & 6.06 & 20.04 \\
  8.2  & 6.24 & 20.75 & 6.16 & 20.41 \\
  8.4  & 6.34 & 21.16 & 6.26 & 20.83 \\
  8.6  & 6.46 & 21.61 & 6.38 & 21.30 \\
  8.8  & 6.58 & 22.12 & 6.50 & 21.81 \\
  9.0  & 6.71 & 22.68 & 6.62 & 22.37 \\
  9.2  & 6.82 & 23.36 & 6.73 & 23.04 \\
  9.4  & 6.90 & 24.10 & 6.81 & 23.77 \\
  9.6  & 6.96 & 24.82 & 6.86 & 24.47 \\
  9.8  & 6.98 & 25.53 & 6.89 & 25.21 \\
\hline
\end{tabular}
\end{center}
\end{table}

\begin{table}
\caption[]{Same as in Table~\ref{tab:TbTs12} 
but for $\rho_\mathrm{b}=4\times10^{11}$~\gcc.}
\label{tab:TbTs_deep}
\begin{center}
\begin{tabular}{ccccc}
\hline
\hline
$\lg T_\mathrm{b}$ &
 $\lg T_\mathrm{sp}$ & $\lg F_\mathrm{bp}$ &
 $\lg T_\mathrm{s}$ & $\lg F_\mathrm{b}$ 
  \\
         & pole  & pole & av  & av 
  \\
\hline
  6.8  &   5.37  &  17.32 &  5.31  &  17.05 \\
  7.0  &   5.49  &  17.77 &  5.42  &  17.49 \\
  7.2  &   5.59  &  18.20 &  5.52  &  17.91 \\
  7.4  &   5.70  &  18.61 &  5.62  &  18.31 \\
  7.6  &   5.80  &  19.01 &  5.72  &  18.70 \\
  7.8  &   5.89  &  19.40 &  5.82  &  19.10 \\
  8.0  &   5.99  &  19.80 &  5.92  &  19.50 \\
  8.2  &   6.09  &  20.20 &  6.02  &  19.91 \\
  8.4  &   6.19  &  20.60 &  6.13  &  20.33 \\
  8.6  &   6.30  &  21.02 &  6.23  &  20.78 \\
  8.8  &   6.40  &  21.53 &  6.34  &  21.39 \\
  9.0  &   6.47  &  22.46 &  6.43  &  22.45 \\
  9.2  &   6.50  &  23.64 &  6.46  &  23.64 \\
  9.4  &   6.51  &  24.76 &  6.46  &  24.77 \\
  9.6  &   6.51  &  25.67 &  6.46  &  25.67 \\
  9.8  &   6.51  &  26.41 &  6.46  &  26.41 \\
\hline
\end{tabular}
\end{center}
\end{table}

\begin{figure}
\epsfysize=72mm
\epsffile{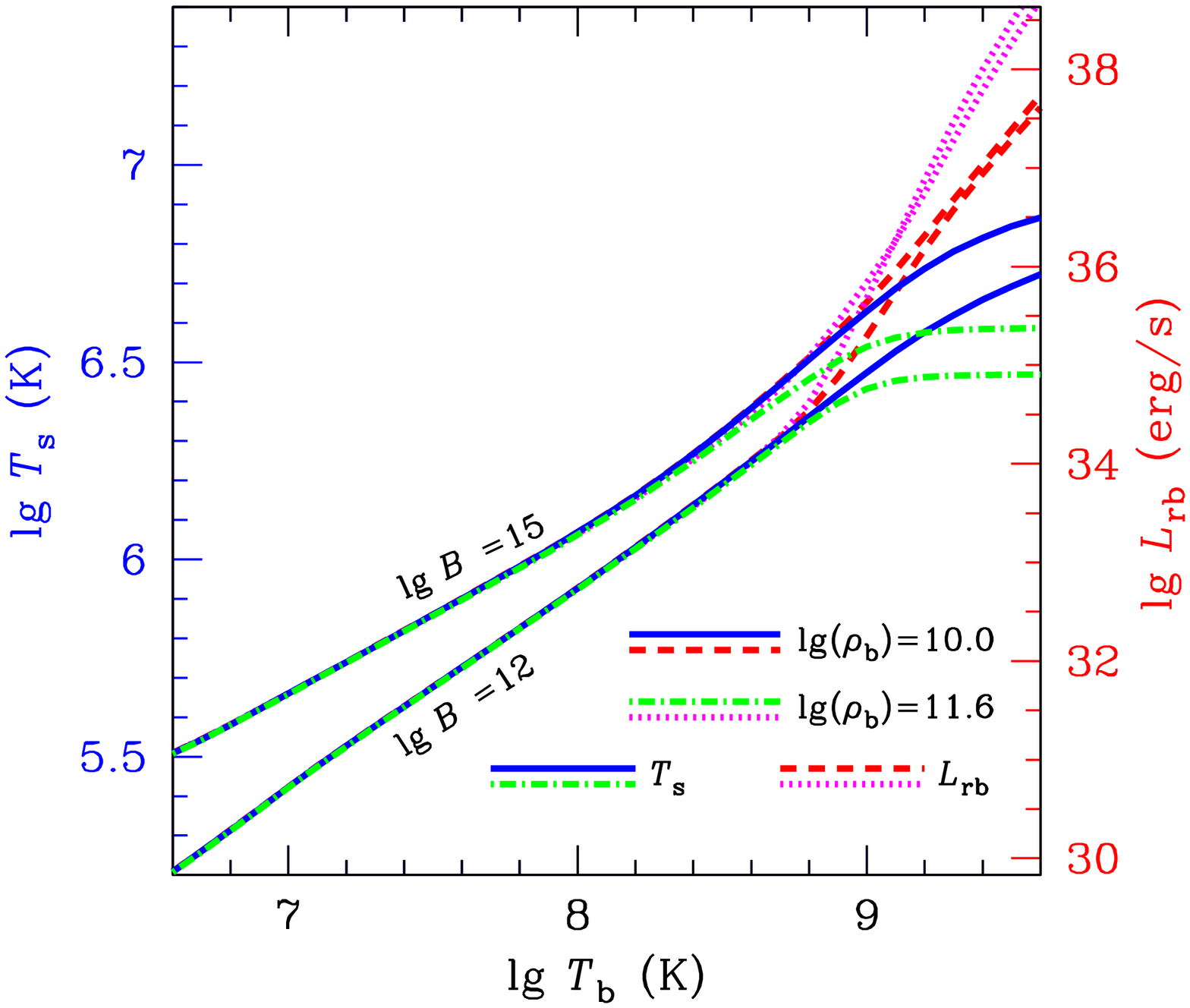}
\caption{
The surface-averaged effective temperature $T_\mathrm{s}$
(left vertical axis, solid and dot-dashed lines)
and the total heat flux $L_\mathrm{rb}$
through the inner boundary of the Fe blanketing
envelope (right vertical axis, dashed and dotted lines)
in the 1.4~$M_\odot$ star 
for $\rho_\mathrm{b}=10^{10}$ \gcc\ (solid and dashed lines)
and $\rho_\mathrm{b}=4\times10^{11}$ \gcc\ (dot-dashed and dotted lines),
for $B=10^{12}$~G (lower curve of each pair) and $10^{15}$~G
(upper curves).
}
\label{fig:append}
\end{figure}

In Tables \ref{tab:TbTs12}--\ref{tab:TbTs_deep}
we present the relations between the temperature $T_\mathrm{b}$
at the inner boundary of the blanketing envelope
($\rho=\rho_\mathrm{b}$) 
and the non-redshifted effective surface temperature
$T_\mathrm{s}$. In addition, we list the values
of the outward radial heat flux $F_\mathrm{b}$ at $\rho=\rho_\mathrm{b}$. 
We assume no heat sources in the envelope
and consider the Fe envelope in the
1.4~$M_\odot$ neutron star of radius $R=12.27$~km with the
dipole magnetic field ($B_\mathrm{p}=10^{12}$ or $10^{15}$~G
at the magnetic pole). We present local values of
$T_\mathrm{s}$ and $F_\mathrm{b}$ at magnetic poles, as well as
surface-averaged values (av).

Table \ref{tab:TbTs12} refers to $B_\mathrm{p}=10^{12}$~G and
$\rho_\mathrm{b}=10^{10}$~\gcc; Table \ref{tab:TbTs15} to
$B_\mathrm{p}=10^{15}$~G and $\rho_\mathrm{b}=10^{10}$~\gcc;
and Table \ref{tab:TbTs_deep} to $B_\mathrm{p}=10^{12}$~G and 
$\rho_\mathrm{b}=4 \times 10^{11}$~\gcc.

Fig.~\ref{fig:append} compares these $T_\mathrm{b}$\,--\,$T_\mathrm{s}$
relations. 
We see that 
$T_\mathrm{s}$ saturates at $T_\mathrm{b}\gtrsim10^9$~K
if $\rho_\mathrm{b}$ is placed at the bottom of the outer crust,
$\rho_\mathrm{b}=4\times10^{11}$ \gcc,
but it does not saturate at lower $\rho_\mathrm{b}=10^{10}$ \gcc.
In the former case the saturation occurs because of the strong neutrino
emission at $\rho_\mathrm{b} \gtrsim 10^{10}$ \gcc\ and high temperatures.
We also plot the total outward heat flux through the
boundary of the blanketing
envelope, $L_\mathrm{rb}$, in such a scale that corresponding 
curves match the $T_\mathrm{b}$\,--\,$T_\mathrm{s}$ 
curves at low $T_\mathrm{b}$.
Then, the deviation of the $T_\mathrm{s}$-curves from the 
$L_\mathrm{rb}$-curves directly measures 
the total energy loss due to neutrino emission in the blanketing envelope
(at $\rho < \rho_\mathrm{b}$).

\label{lastpage}

\end{document}